\documentclass[twocolumn]{jpsj3}
\usepackage{txfonts}
\usepackage{bm}
\usepackage{amsmath}
\usepackage{color,multirow}
\usepackage{graphicx}
\title{Bulk Electronic State of Superconducting Topological Insulator}

\author{Tatsuki Hashimoto, Keiji Yada, Ai Yamakage, Masatoshi Sato and Yukio Tanaka}
\inst{Department of Applied Physics, Nagoya University, Nagoya 464-8603, Japan}

\abst{We study the electronic properties of a superconducting topological
 insulator whose parent material is a topological insulator. 
We calculate the temperature dependence of the specific heat and spin
 susceptibility for four promising superconducting pairings proposed by
 L. Fu and E. Berg [Phys. Rev. Lett. {\bf105} (2010) 097001].  
Since the line shapes of the temperature dependence of specific heat are almost
identical among three of the four pairings, 
it is difficult to identify them simply from the
specific heat. 
On the other hand,
we obtain
wide variations of the temperature dependence of spin susceptibility for
 each pairing, reflecting the spin structure of the Cooper pair. 
We propose that the pairing symmetry of a superconducting topological
 insulator can be determined from measurement of the Knight shift by
 changing the direction of the applied magnetic field. }

\kword{superconducting topological insulator, topological superconductor, topological insulator, unconventional superconductivity, odd-parity, spin-orbit interaction, multi-band system, spin susceptibility, specific heat}

\begin{document}
\maketitle

\section{Introduction}

Topological insulators (TIs) are a newly discovered state of matter supporting massless Dirac fermions on the surface and characterized by nonzero topological numbers defined in the bulk.\cite{Hasan,XLQi}
Because of the presence of the surface Dirac fermions, 
TIs have the potential to exhibit rich transport and electromagnetic response properties, which may be applicable for future devices.
The superconducting analog of TIs are topological superconductors,\cite{Tanaka1,XLQi,Schnyder,Sato09,Sato10} which have Majorana fermions\cite{Wilczek} on the surface as Andreev bound states (ABSs).
In these materials, topological invariants can be defined in the bulk Hamiltonian.
There are several types of topological superconductors, e.g., the chiral $p$-wave superconducting state in Sr$_2$RuO$_4$\cite{Maeno,Mackenzie,Furusaki,Stone,Kashiwaya}and the helical superconducting state realized in non-centrosymmetric superconductors.\cite{Tanaka,Sato1}
The realization of a topological superconductor is of particular interest from the viewpoint of quantum devices and quantum computations.\cite{Sato1,Tanaka,Sato2,Sato3,Sato4,Sau,Alicea1,Lutchyn1,Oreg,Lutchyn2,Alicea2,FuKane1,FuKane2,Akhmerov,Law,Tanaka2,Linder,Yamakage1,Yamakage2,Beenackker}

Recently, 
the carrier-doped TI Cu$_x$Bi$_2$Se$_3$ has been revealed to be a superconductor.\cite{Hor}
Hereafter, 
we refer to a superconductor based on a TI as a superconducting topological insulator (STI).
In tunneling spectroscopy,\cite{Sasaki} Cu$_x$Bi$_2$Se$_3$ shows a zero-bias conductance peak (ZBCP).
This means that Cu$_x$Bi$_2$Se$_3$ can be regarded as a topological
superconductor since the ZBCP signifies the existence of gapless ABSs
\cite{Hu,Tanaka95,Kashiwaya00} on the surface,  
which is a direct consequence of topological superconductivity.
Interestingly, it has been clarified that an STI supports anomalous
ABSs different from those of other topological
superconductors, \cite{HaoLee,Hsieh,Yamakage1} 
and the resulting transport property also becomes
anomalous.\cite{Hsieh,Yamakage1}
In this sense, STIs are a new type of topological superconductor, and have
attracted much interest.
Moreover, there are several experimental results supporting the generation of an STI by the proximity effect, \cite{Koren1,Koren2} while a recent study based on scanning tunneling spectroscopy has reported conventional superconductivity in an STI.\cite{Stroscio}

There have been many relevant studies on Cu$_x$Bi$_2$Se$_3$.\cite{Hor,Wray1,Wray2,Kriener1,Kriener2,Sasaki,Das,Kriener4,Nagai}
However, up to now, the symmetry of the superconductivity of Cu$_x$Bi$_2$Se$_3$  still remains unknown, while its topological properties crucially depend on the pairing symmetry. Although the specific heat has been measured, it is difficult to establish the superconducting symmetry only from the data of specific heat.
More careful analysis with the help of microscopic calculations is needed.
In order to clarify the superconducting symmetry, it is useful to
analyze the spin susceptibility in addition to the specific heat since the spin
susceptibility is directly related to the spin structure of the superconducting
pairing. Indeed, to determine the pairing symmetry of unconventional
superconductors such as cuprates, Sr$_2$RuO$_4$ and UPt$_3$, the measurement of specific heat and spin susceptibility has played an important role.\cite{Scalapino,RevModPhys.67.503,RevModPhys.72.969,Maeno,Mackenzie,NomuraYamada,PhysRevLett.87.057001,PhysRevLett.80.3129,PhysRevB.77.184515}

In this paper, we clarify the temperature dependence of specific heat
and spin susceptibility for the possible superconducting pairings.
In contrast to unconventional superconductors, because of the
strong spin-orbit interaction, a mixture of orbital degrees of freedom is essential to realize
unconventional superconductivity in an STI. 
Therefore, a careful analysis is needed to study the specific heat and
spin susceptibility. 
Actually, we find that the quasi-particle spectra of an STI are very
different from those of the previously studied unconventional
superconductors, and thus the spin susceptibility depends on the $d$-vector
nontrivially. In particular, even for a spin-singlet superconducting gap
($\Delta_3$ in the text),
an STI may show $T$-independent spin susceptibility.  
On the basis of the non trivial behaviors of the specific heat and spin
susceptibility, it is possible to determine the pairing symmetry in an
STI.  

The paper is organized as follows.
In \S\ref{model}, we give the model Hamiltonian of an STI and the energy spectra for the possible superconducting pairings.
The numerical results and discussion on the temperature dependences of the specific heat and spin susceptibility are given in \S\ref{specificheat} and \S\ref{susceptibility}, respectively.
We compare our results of specific heat with the experimental data.\cite{Kriener1}
In \S\ref{summary}, we summarize our results and propose how to experimentally determine the superconducting symmetry of an STI.

\section{Model}
\label{model}
\begin{figure}
\begin{center}
\includegraphics[width=5cm]{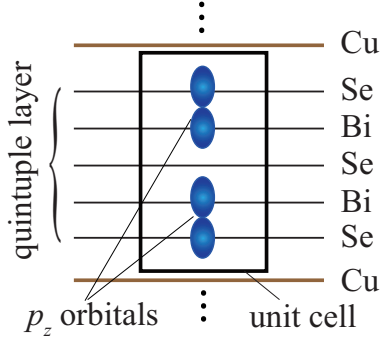}
\caption{
(Color online) Two $p_z$ orbitals in the quintuple layer of Bi$_2$Se$_3$.
}
\label{figorbital}
\end{center}
\end{figure}

For our model of an STI,
we start with the Bogoliubov-de Gennes (BdG) Hamiltonian proposed in ref. \citen{FuBerg},
\begin{eqnarray}
H({\bm k})=H_0({\bm k})\tau_z+\Delta_\ell\tau_x,
\label{hamiltonian}
\end{eqnarray}
where $\ell(=1,2,3,4)$ represents the type of pair potential.
The normal part of the Hamiltonian $H_0({\bm k})$ is the low-energy effective model of a topological insulator based on ${\bm k}\cdot{\bm p}$ theory given by
\begin{eqnarray}
H_0({\bm k})&=&c({\bm k})+m({\bm k})\sigma_x\nonumber \\ &\ &\hspace{4em}+v_zk_z\sigma_y +v(k_xs_y-k_ys_x)\sigma_z,
\label{h00} \\ 
m({\bm k})&=&m_0+m_1k_z^2+m_2(k_x^2+k_y^2),
\label{m} \\
c({\bm k})&=&-\mu+c_1k_z^2+c_2(k_x^2+k_y^2).
\label{c}
\end{eqnarray}
$s_i$, $\sigma_i$ and $\tau_i$ $(i=x,y,z)$ are the Pauli matrices in the spin, orbital and Nambu spaces respectively. ]
The basis of the orbitals consists of effective $p_z$ orbitals constituted from the $p_z$ orbitals of Se and Bi on the upper and lower sides of the quintuple layer, as shown in Fig. \ref{figorbital}.
Hereafter, we call this basis the ``orbital basis".
On the other hand, we refer to the basis diagonalizing $H_0(\bm k)$ as the ``band basis", which is introduced in \S\ref{bandbase}. 
In this model, the normal part $H_0({\bm k})$ is equivalent to the model proposed in refs. \citen{Zhang} and \citen{Liu} under the unitary transformation.
In the following, we use the tight-binding model, which is equivalent to the above model at low energy.
We consider a hexagonal lattice where two-dimensional triangular lattices are stacked along the $c$-axis.\cite{HaoLee,Sasaki}
Then, the tight-binding Hamiltonian is obtained by the following substitution in the ${\bm k}\cdot{\bm p}$ Hamiltonian given by eqs. (\ref{h00})-(\ref{c}).
\begin{align}
k_x&\rightarrow \frac{2}{\sqrt3a}\sin\frac{\sqrt3k_xa}{2}\cos\frac{k_ya}{2}\equiv f_x({\bm k}),\\
k_y&\rightarrow \frac{2}{3a}(\cos\frac{\sqrt3k_xa}{2}\sin\frac{k_ya}{2}+\sin k_ya)\equiv f_y({\bm k}),\\
k_z&\rightarrow \frac{1}{c}\sin k_zc\equiv f_z({\bm k}),\\
k_z^2&\rightarrow \frac{2}{c^2}(1-\cos k_zc)\equiv f_\perp({\bm k}),\\
k_x^2+k_y^2&\rightarrow \frac{4}{3a^2}\large{(}3-2\cos \frac{\sqrt3k_xa}{2}\cos\frac{k_ya}{2}- \cos k_ya\large{)} \nonumber \\ &\equiv f_\parallel({\bm k}),
\end{align}
where $a$ and $c$ are the lattice constants.
In this hexagonal lattice, the primitive lattice vectors are $(\sqrt3a/2,a/2,0)$, $(0,a,0)$, and $(0,0,c)$,
although the actual crystal structure is not hexagonal but rhombohedral.\cite{Zhang,Liu}
This simplification does not affect the low-energy excitations.
Then, the normal part of the Hamiltonian is {summarized} as follows:
\begin{eqnarray}
H_0({\bm k})&=&c({\bm k})+m({\bm k})\sigma_x\notag\\  &\ &+(a_x({\bm k})s_y-a_y({\bm k})s_s)\sigma_z+b({\bm k})\sigma_y,\label{h0}\\
c({\bm k})&=&-\mu+c_1 f_\perp({\bm k})+c_2 f_\parallel({\bm k}),\\
m({\bm k})&=&m_0+m_1 f_\perp({\bm k})+m_2 f_\parallel({\bm k}),\\
a_{x,y}({\bm k})&=&v f_{x,y}({\bm k}),\\
b({\bm k})&=&v_z f_{z}({\bm k}).
\end{eqnarray}
Here, we choose the chemical potential $\mu=0.5$ eV, since the chemical potential measured from the surface Dirac point
is 0.4-0.5 eV according to ref. \citen{Wray1}.
We use the values of the parameters {$c_2$, $m_0$, $m_2$ and $v$} as given in ref. \citen{Liu}.
On the other hand, for $c_1$, $m_1$ and $v_z$, we choose the different values given in ref. \citen{Liu}, which involve hopping along the $c$-axis.
Since the parameterization performed in ref. \citen{Liu} is based on the dispersion around the $\Gamma$-point,
the difference in the dispersion near the zone boundary between the first-principles calculation in ref. \citen{Liu} and our tight-binding model is considerably large.
However, the Fermi surface becomes cylindrical if we use the same parameters given in ref. \citen{Liu} although the correct shape of the Fermi surface is an spheroidal one.
Thus, we choose the values of {$c_1$, $m_1$ and $v_z$ as $c_1/c^2=0.024$, $m_1/c^2=0.216$ and $v_z/c=0.32$ (eV)} to fit the energy dispersion
for the $\Gamma$-Z direction obtained in ref. \citen{Liu}.
These parameters give the spheroidal Fermi surface consistent with the first-principles calculation.
This parameterization is crucial
since the specific heat and spin susceptibility in actual Cu$_x$Bi$_2$Se$_3$
cannot be reproduced
if we use a cylindrical Fermi surface.
In addition, to obtain the topological superconductivity in three
dimensions, the correct Fermi surface topology is needed.\cite{Sato09,
Sato10,FuBerg}

Next, we consider the pair potentials.
We assume that each pair potential is independent of momentum since
the present material is not a strongly correlated system.\cite{FuBerg}
In this case, 
the pair potentials are classified into four types of irreducible representation for the $D_{3d}$ point group.
The matrix forms of the pairings $\Delta_{1}$, $\Delta_{2}$, $\Delta_{3}$, and $\Delta_{4}$ are shown in the first column of Table \ref{table1}.
$\Delta_{1}$ and $\Delta_{3}$ are spin-singlet intra-orbital pairings, whereas $\Delta_{2}$ and $\Delta_{4}$ are spin-triplet inter-orbital pairings in the orbital basis. 
Note that intra-orbital repulsion can be relevant to inter-orbital pairings even though this system is not a strongly correlated system.

We diagonalize the BdG Hamiltonian [eq. (\ref{hamiltonian})]. We obtain four branches of the bulk spectrum $E_{\gamma}$
($\gamma=1,2,3,4$) for each pairing, 
\begin{eqnarray}
E_1({\bm k})&=&\sqrt{\xi^2({\bm
 k})+2\sqrt{\eta^2({\bm k})c^2({\bm k})+\zeta^2({\bm
 k})\Delta^2}},
\\
E_2({\bm k})&=&\sqrt{\xi^2({\bm
 k})-2\sqrt{\eta^2({\bm k})c^2({\bm k})+\zeta^2({\bm
 k})\Delta^2}},
\\
E_3({\bm k})&=&-\sqrt{\xi^2({\bm
 k})+2\sqrt{\eta^2({\bm k})c^2({\bm k})+\zeta^2({\bm
 k})\Delta^2}},
\\
E_4({\bm k})&=&-\sqrt{\xi^2({\bm
 k})-2\sqrt{\eta^2({\bm k})c^2({\bm k})+\zeta^2({\bm
 k})\Delta^2}},
\end{eqnarray}
with
\begin{eqnarray}
\eta^2({\bm k})&=&m^2({\bm k})+a_x^2({\bm k})+a_y^2({\bm k})+b^2({\bm k}),\\
\xi^2({\bm k})&=&\eta^2({\bm k})+c^2({\bm k})+\Delta^2.
\end{eqnarray}
The difference in the energy gap structure in each pairing originates from $\zeta^2({\bm k})$,
\begin{eqnarray}
\Delta_{1}:\zeta^2({\bm k})&=&0,\\
\Delta_{2}:\zeta^2({\bm k})&=&m^2({\bm k}),\\
\Delta_{3}:\zeta^2({\bm k})&=&m^2({\bm k})+b^2({\bm k}),\\
\Delta_{4}:\zeta^2({\bm k})&=&m^2({\bm k})+a_y^2({\bm k}).
\end{eqnarray}
 The energy gap structure of $\Delta_{1}$ is an isotropic full gap, which is the same as that of conventional BCS superconductors. 
In other cases, because of the presence of $\zeta$, the energy gap is modified from the BCS gap structure.
$\Delta_{2}$ is an anisotropic full-gap pairing. 
In the cases of $\Delta_{3}$ and $\Delta_{4}$, the energy gap has point nodes. The point nodes for $\Delta_{3}$ are on the poles. 
In the case of $\Delta_{4}$, point nodes appear on the $k_y$-axis.
Although,
in general,
$\Delta_4$ is a linear combination of $\Delta \sigma_y s_x$ and $\Delta \sigma_y s_y$, we can choose $\Delta_4 = \Delta\sigma_y s_x$ without loss of a generality.
The energy gap of $E_{\gamma}$ is influenced by the spin-orbit interaction $v$.
To elucidate the role of the spin-orbit interaction,
we also consider the case of $v=0$.
In this case,
$E_\gamma$ for $\Delta_1$ has a full gap, $E_{\gamma}$ for $\Delta_2$ and $\Delta_4$ have line nodes on the equator, 
and $E_\gamma$ for $\Delta_3$ is gapless.

\begin{table}
\begin{tabular}{l|c|c|c|c}
\hline\hline
pair potential                 & rep. & spin & orbital & energy gap\\
\hline
$\Delta_{1}=\Delta$            & $A_{1g}$  & singlet & intra & isotropic full gap\\
 & & & & (isotropic full gap)
\\
$\Delta_{2}=\Delta \sigma_y s_z$& $A_{1u}$  & triplet & inter & anisotropic full gap\\
 & & & & (line node on equator)
\\
$\Delta_{3}=\Delta\sigma_z$    & $A_{2u}$  & singlet & intra & point nodes at poles\\
 & & & & (gapless)
\\
$\Delta_{4}=\Delta \sigma_y s_x$& $E_{u}$   & triplet & inter & point node on equator \\
 & & & & (line node on equator)
\\
\hline\hline
\end{tabular}
\caption{Irreducible representation, spin state, orbital state and, energy gap structure in each pairing symmetry.
In the brackets we denote the gap structure for $v=0$.
}
{\label{table1}}
\end{table}

\section{Specific Heat}
\label{specificheat}
In this section, we calculate the specific heat below $T_c$ for each pairing symmetry.
The specific heat is given by

\begin{eqnarray}
C_{s}&=&-\frac{2\beta}{N}\sum_{\bm k \gamma}\left(-\frac{\partial f(E_{\gamma}({\bm k}))}{\partial E_{\gamma}({\bm k})}\right)\left(E_{\gamma}^2({\bm k})+\frac{\beta}{2}\frac{\partial E^2_{\gamma}({\bm k})}{\partial \beta}\right)\nonumber\\
&=&-\frac{2\beta}{N}\sum_{\bm k \gamma}\left(-\frac{\partial f(E_{\gamma}({\bm k}))}{\partial E_{\gamma}({\bm k})}\right)\nonumber\\ &\ &\hspace{4em}\times\left(E_{\gamma}^2({\bm k})+\beta E_{\gamma}({\bm k})\frac{\partial \Delta}{\partial \beta}\frac{\partial E_{\gamma}({\bm k})}{\partial \Delta}\right),
\end{eqnarray}
where $N$ is the number of unit cells and $\beta$ is $1/k_{\rm B}T$, with the Boltzmann constant $k_B$ and temperature $T$.
We assume that the temperature dependence of the pairing potential is the scaled BCS one, $\Delta(T)=(\alpha/\alpha_{\rm{BCS}})\Delta_{\rm{BCS}}(T)$.
The magnitude of $\alpha$ gives the ratio of $\Delta(T=0)$ to $T_c$, i.e., $\alpha=\Delta(T=0)/(k_BT_c)$.
This model is known as the $\alpha$-model.\cite{Padamsee}
For $\Delta_{\rm{BCS}}(T)$, we use the following phenomenological form:\cite{BM}
\begin{eqnarray}
\Delta_{\rm{BCS}}(T)=\alpha_{\rm{BCS}}k_BT_c\tanh\left(1.74\sqrt{\frac{T_c}{T}-1}\right),
\end{eqnarray}
with $\alpha_{\rm{BCS}}=1.76$.

Since $\alpha$ is a material-dependent parameter and it often deviates from the BCS value $\alpha_{\rm{BCS}}=1.76$,
we use two different values.
One is $\alpha=\alpha_{\rm{BCS}}$ and the other is $\alpha=\alpha_0$, where $C_s(T)/T$ at $T=0.53T_{\rm c}  \equiv T_0$ becomes equal to that for the normal state $C_n(T)/T\simeq C_n(T_c)/T_c$
as observed in specific heat measurements.\cite{Kriener1}

\subsection{Isotropic full gap $\Delta_1$}
\begin{figure}[htbp]
\begin{center}
\includegraphics[width=7.5cm]{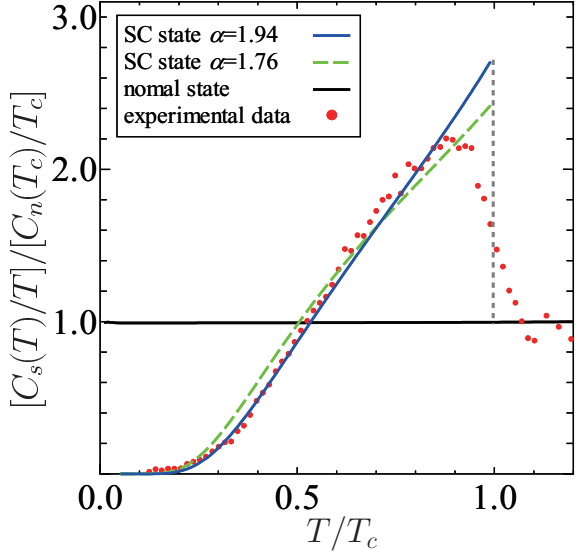}
\caption{(Color online) Temperature dependence of specific heat for $\Delta_1$ with $\alpha=1.94$ (blue solid line) and $\alpha=1.76$ (green dashed line). The black solid line shows the specific heat for the normal state. Red dotted circles show the experimental data in ref. \citen{Kriener1}}\label{c1}.
\end{center}
\end{figure}
In Fig. \ref{c1}, we show the temperature dependence of $C_s$ for $\alpha=\alpha_0=1.94$ (blue solid line) and $\alpha=1.76$ (green dashed line).
In the case of $\Delta_1$, the energy spectrum is given by $E({\bm k})=\pm\sqrt{\varepsilon_\pm^2({\bm k})+\Delta^2}$,
where $\varepsilon_\pm^2({\bm k})$ is the dispersion of the normal state.
Therefore, the energy gap structure becomes an isotropic $s$-wave one.
Thus, the specific heat near $T=0$ shows exponential behavior.
If we choose $\alpha = \alpha_{\rm BCS}$, $T_0$ and the magnitude of the specific-heat jump are smaller than those obtained experimentally.
To fit the experimental data,
we choose $\alpha=\alpha_0=1.94$.
Then,
to satisfy the entropy balance relation,
\begin{eqnarray}
\int_0^{T_c}dT\frac{C_s(T)-C_n(T)}{T}=0,
\end{eqnarray}
the magnitude of the specific-heat jump at $\alpha=1.94$ becomes larger than that for $\alpha = \alpha_{\rm BCS}$.

In the case of $\alpha=\alpha_0$, the magnitude of the specific heat jump and the line shape are similar to those of the experimental ones.
Note that the analysis performed in ref. \citen{Kriener1} is based on an isotropic $s$-wave gap and therefore the obtained value of $\alpha_0$ is almost the same.
On the other hand, the value of $\alpha$ can also be estimated from 
the upper and lower critical field in ref. \citen{Kriener1}.
The estimated value is $\alpha=2.3\equiv\alpha_c$.
Therefore, the value of $\alpha_0$ for $\Delta_1$ deviates from that of
$\alpha_c$.
However, if we add a small ${\bm k}$-dependent term allowed in the $A_{1g}$
representation to $\Delta_1$,
then the magnitude of the specific heat jump for $\alpha=\alpha_{\rm{BCS}}$
can be small, the values of $\alpha_0$ become large and 
$\alpha_0=2.3$ might be obtained.

\subsection{Anisotropic full gap $\Delta_2$}
\begin{figure}[htbp]
\begin{center}
\includegraphics[width=7.5cm]{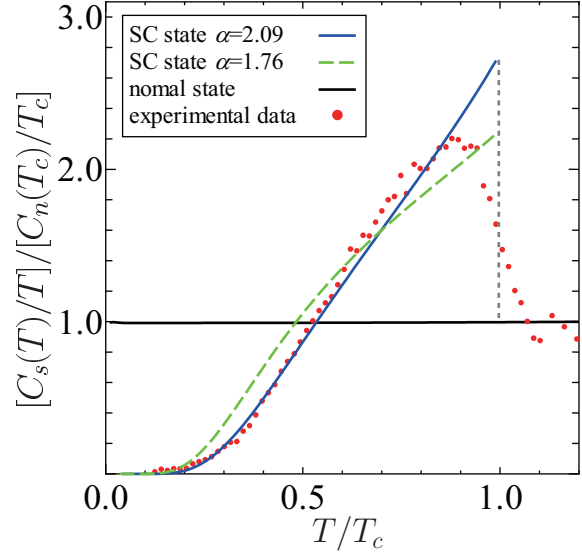}
\caption{(Color online) Temperature dependence of specific heat for $\Delta_2$ with $\alpha=2.09$. The dotted circles show the experimental data in ref. \citen{Kriener1}.}\label{c2}
\end{center}
\end{figure}
In Fig. \ref{c2}, we show the temperature dependence of $C_s(T)$ for $\alpha=\alpha_0=2.09$ (blue solid line) and $\alpha=1.76$ (green dashed line).
Since the energy gap structure is fully gapped, the exponential behavior appears near $T=0$ as in the case of $\Delta_1$.
On the other hand, the magnitude of the specific heat jump for $\alpha=\alpha_{\rm{BCS}}$ is smaller than that for $\Delta_1$ owing to the anisotropy of the energy gap.
Therefore, to reproduce the experimental data, we need a larger value of $\alpha_0$ than for the case of $\Delta_1$, $\alpha_0=2.09$.
This value is closer to $\alpha_c=2.3$ than that for $\Delta_1$.
The magnitude of the specific heat jump and the line shape for $\alpha=\alpha_0$ are similar to those of the experimental ones.

\subsection{Point nodes at polar $\Delta_3$}
\label{sped3}
\begin{figure}[htbp]
\begin{center}
\includegraphics[width=7.5cm]{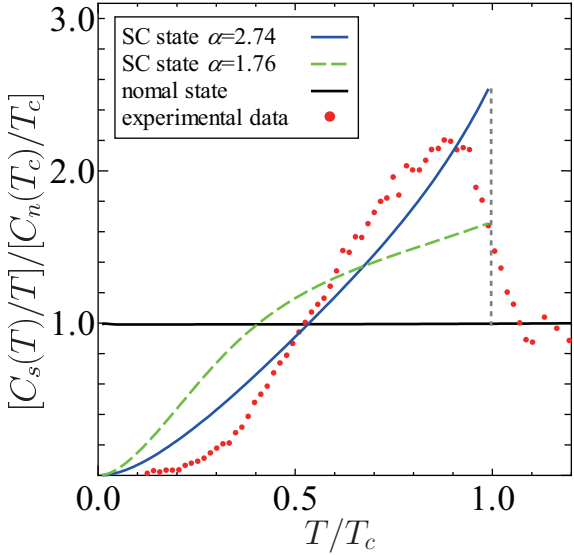}
\caption{(Color online) Temperature dependence of specific heat for $\Delta_3$ with $\alpha=2.74$. The dotted circles show the experimental data in ref. \citen{Kriener1}.}\label{c3}
\end{center}
\end{figure}

In Fig. \ref{c3}, we show the temperature dependence of $C_s(T)$ for $\alpha=\alpha_0=2.74$ (blue solid line) and $\alpha=1.76$ (green dashed line).
In the case of $\Delta_3$, the energy dispersion has point nodes along the $k_z$-axis.
Thus, $C_s(T)/T$ has $T^2$-behavior near $T=0$.
The magnitude of the jump for $\alpha=\alpha_{\rm{BCS}}$ is the smallest among the four pair potentials considered in this paper.
This small jump originates from the gapless nature of this pair potential.
In the case of $v=0$, the energy dispersion for $\Delta_3$ is given by $E({\bm k})=\pm \sqrt{m^2({\bm k})+b^2({\bm k})}\pm\sqrt{c({\bm k})^2+\Delta^2}$.
This energy spectrum becomes gapless when $m^2({\bm k})+b^2({\bm k})=c({\bm k})^2+\Delta^2$:
The parameters of an STI  satisfy the following relations.
\begin{eqnarray}
m_0^2-\mu^2-\Delta^2&<&0,\\
m_1^2-c_1^2&>&0,\\
m_2^2-c_2^2&>&0.
\end{eqnarray}
In this case, the energy spectrum becomes gapless near the Fermi surface in any direction of ${\bm k}$.
Thus, $C_s(T)/T$ with $v=0$ is $T$-independent.
In the presence of the spin-orbit interaction, these gapless energy spectra still remain in the direction of the $k_z$-axis, and point nodes are formed
since $a_{x,y}({\bm k})=0$ in this direction.
In directions other than $k_z$, the energy gap is
generated by the spin-orbit coupling $v$, but the gap is 
smaller than those of the other pairings.
Therefore, the $T$-dependence of $C_s(T)/T$ for
$\alpha=\alpha_{\rm{BCS}}$ remains relatively small.
This is the reason why the specific heat jump is small for $\Delta_3$ compared with that for the other pair potentials.
If we use $\alpha=\alpha_0=2.74$, we can make the specific heat jump similar to the experimental one and $[C_s(T_0)/T_0]/[C_n(T_c)/T_c]$ becomes equal to unity.
However, the line shape does not reproduce the experimental data.
In addition, the value of $\alpha_0$ is much larger than the experimental value of $\alpha_c=2.3$.

\subsection{Point nodes on equator $\Delta_4$}
\begin{figure}[htbp]
\begin{center}
\includegraphics[width=7.5cm]{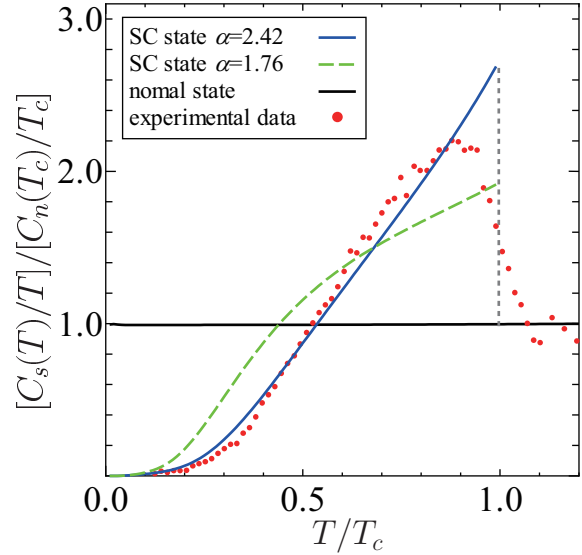}
\caption{(Color online) Temperature dependence of specific heat for $\Delta_4$ with $\alpha=2.42$. The dotted circles show the experimental data in ref. \citen{Kriener1}.}\label{c4}
\end{center}
\end{figure}

In Fig. \ref{c4}, we show the temperature dependence of $C_s(T)$ for $\alpha=\alpha_0=2.42$ (blue solid line) and $\alpha=1.76$ (green dashed line).
In the case of $\Delta_4$, the energy spectrum has point nodes along the $k_y$-axis.
Therefore, $C_s(T)/T$ has $T^2$-behavior near $T=0$ as in the case of $\Delta_3$.
However, in the case of $\Delta_4$, the energy spectrum does not become gapless even when the spin-orbit interaction is absent.
Thus, the coefficient of $T^2$ is smaller than that in the case of $\Delta_3$ for $\alpha=\alpha_{\rm{BCS}}$,
and the magnitude of the specific heat jump is larger than that for $\Delta_3$.
As a result, the line shape with $\alpha=\alpha_0=2.42$ is considerably closer to the experimental one than in the case of $\Delta_3$.
The obtained value of $\alpha=2.42$ is the closest to the experimental one, $\alpha_c$, among the four types of pairing symmetry considered in this paper.

Here, we summarize the results of the specific heat.
We have calculated the specific heat for $\alpha=\alpha_{\rm{BCS}}$ and $\alpha_0$ in each pairing symmetry.
For $\alpha=\alpha_0$, we obtain line shapes similar to the experimental one in the cases of $\Delta_1$, $\Delta_2$, and $\Delta_4$.
The obtained values of $\alpha_0$ are $\alpha=1.94$, 2.09, 2.74, and 2.42 for $\Delta_1$, $\Delta_2$, $\Delta_3$, and $\Delta_4$, respectively.
The values of $\alpha$ for $\Delta_2$ and $\Delta_4$ are closer to the experimental one ($\alpha_c=2.3$), than for the other pair potentials.

\section{Spin Susceptibility}
\label{susceptibility}
From the temperature dependence of spin susceptibility, 
one can determine the spin structure of Cooper pairs.
Namely, for a spin-singlet superconductor, the spin susceptibility along any direction decreases with decreasing $T$for $T < T_{\rm c}$ and vanishes at $T=0$.
On the other hand, for a spin-triplet superconductor, 
only the spin susceptibility parallel to the direction of the $d$-vector decreases with decreasing $T$ and vanishes at $T=0$, and the spin susceptibility perpendicular to the $d$-vector is independent of $T$.
However, the temperature dependence of the spin susceptibility of an STI is not simple because spin-singlet and spin-triplet components can mix in the band basis owing to the dependence of the spin-orbit interaction on the pair potential.

Nevertheless, we show here that it is possible to determine the spin structure of an STI, even if the spin-orbit interaction is present.
The temperature dependences of spin susceptibility for each possible pairing are different.
For $\Delta_1$, the spin susceptibility along any direction decreases as temperature decreases  
since $\Delta_1$ is a spin-singlet pair potential in the band basis.
On the other hand, that along the $z$-axis for $\Delta_3$ is independent of temperature, although those along the $x$- and $y$-axes decrease with decreasing temperature.
Spin susceptibilities with $\Delta_2$ and $\Delta_4$ along the $d$-vector ($\bm d \parallel z$ for $\Delta_2$ and $\bm d \parallel x$ for $\Delta_4$) decrease with decreasing temperature.
Those perpendicular to the $d$-vector are almost independent of temperature.

We now comment on the effects of the spin-orbit interaction on spin susceptibility.
There are three effects.
The first one is the Van Vleck susceptibility, which originates from inter-band (off-diagonal) matrix elements.
The Van Vleck susceptibility can appear in multiband systems with the spin-orbit interaction. 
This leads to a non zero value of spin susceptibility at $T=0$ (see Appendix \ref{vanvleck}).
The second one is the rotation of the $d$-vector by the unitary transformation from the orbital basis to the band basis, after which the $d$-vector in the band basis is not parallel to the Zeeman magnetic field, even when the $d$-vector in the orbital basis is.
This also induces a non zero value of spin susceptibility at $T=0$.
Additionally, 
the spin susceptibility perpendicular to the $d$-vector in the orbital basis also decreases slightly with decreasing temperature for $T<T_{\rm c}$.
This behavior occurs in the case of $\Delta_2$ and $\Delta_4$.
The third one is the generation of spin-singlet and spin-triplet pair potentials in the band basis from those in the orbital basis, respectively. We summarize these effects for each pairing in Table \ref{summary_sus}.
In the following sections, we shall discuss the temperature dependence of the spin susceptibility in each pairing.

\begin{table}
\centering

\begin{tabular}{l|c}
 \hline\hline
 pair potential &  effects of SOI 
\\
 \hline
 {$\Delta_1 = \Delta$} & Van Vleck 
 \\
 \hline
 \multirow{2}{*}{$\Delta_2 = \Delta \sigma_y s_z$} & Van Vleck 
 \\
   & rotation of $d$-vector
 \\
 \hline
 \multirow{2}{*}{$\Delta_3 = \Delta \sigma_z$} & Van Vleck 
 \\
   & induced spin-triplet
 \\
 \hline
 \multirow{3}{*}{$\Delta_4 = \Delta \sigma_y s_x$} & Van Vleck 
 \\
  & rotation of $d$-vector
 \\
  & induced spin-singlet
 \\
 \hline\hline
\end{tabular}
\caption{Summary of effects of spin-orbit interaction on spin susceptibility.
The effects of the spin-orbit interaction are the Van Vleck susceptibility, rotation of the $d$-vector, and the induction of spin-singlet and spin-triplet pair potentials.
}
\label{summary_sus}
\end{table}

\subsection{Kubo formula for spin susceptibility}

First, we give the Zeeman term in an STI and the Kubo formula for spin susceptibility.
The Zeeman term $H_{\rm Z}(\bm k)$ is given by
\begin{align}
 H_{\rm Z}(\bm k) =  
 \sum_{i=x,y,z} 
\sum_{\mu=0,x,y,z}  h_i \mu_{\rm B}\frac{g_{i \mu}}{2} s_i \sigma_\mu,
\end{align}
where $\mu_{\rm B}$ is the Bohr magneton, $h_i$ is the $i$ th component of the Zeeman field, $g_{i \mu}$ is the $g$-factor of the parent topological insulator, and $\sigma_0$ is the $2 \times 2$ identity matrix in the orbital space.
The spin susceptibility along the $i$-axis is given by the Kubo formula:
\begin{align}
 \chi_i &= -\mu_{\rm B}^2 
\lim_{\bm q \to 0}
\frac{1}{V}\sum_{\bm k\alpha\beta \mu} 
\frac{f(E_\alpha(\bm k)) - f(E_{\beta}(\bm k + \bm q)) }
{E_{\alpha}(\bm k) - E_{\beta}(\bm k + \bm q) + i 0} 
\nonumber\\  & \qquad \times
\langle \alpha | s_i | \beta \rangle
\langle \beta | \frac{g_{i\mu}}{2} s_i \sigma_\mu | \alpha \rangle .
\end{align}
In the actual calculation, we set $g_{i \mu}$ to be those of Bi$_2$Se$_3$:\cite{Liu} $g_{x0}= g_{y0} = -8.92, g_{z0}= -21.3, g_{xx} = g_{yx} = 0.68, g_{zx}=-29.5$. The other $g$-factors are chosen to be zero. The temperature dependence of $\Delta$ is the same as that estimated from the specific heat measurement with $\alpha_c=2.3$.\cite{Kriener1}

\subsection{Spin structure in the band basis}
\label{bandbase}
\begin{figure}
\centering
\includegraphics[scale=0.55]{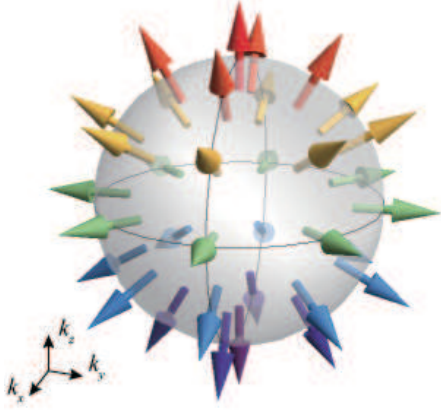}
\caption{(Color online) Vector field plot of the conduction-band  component of $\tilde {\bm d}(\bm k)$ given by eq. \ref{del2d} for $\Delta_2$ [(1,1)-component].
}
\label{d211}
\end{figure}

\begin{figure}
\centering
\includegraphics[scale=0.55]{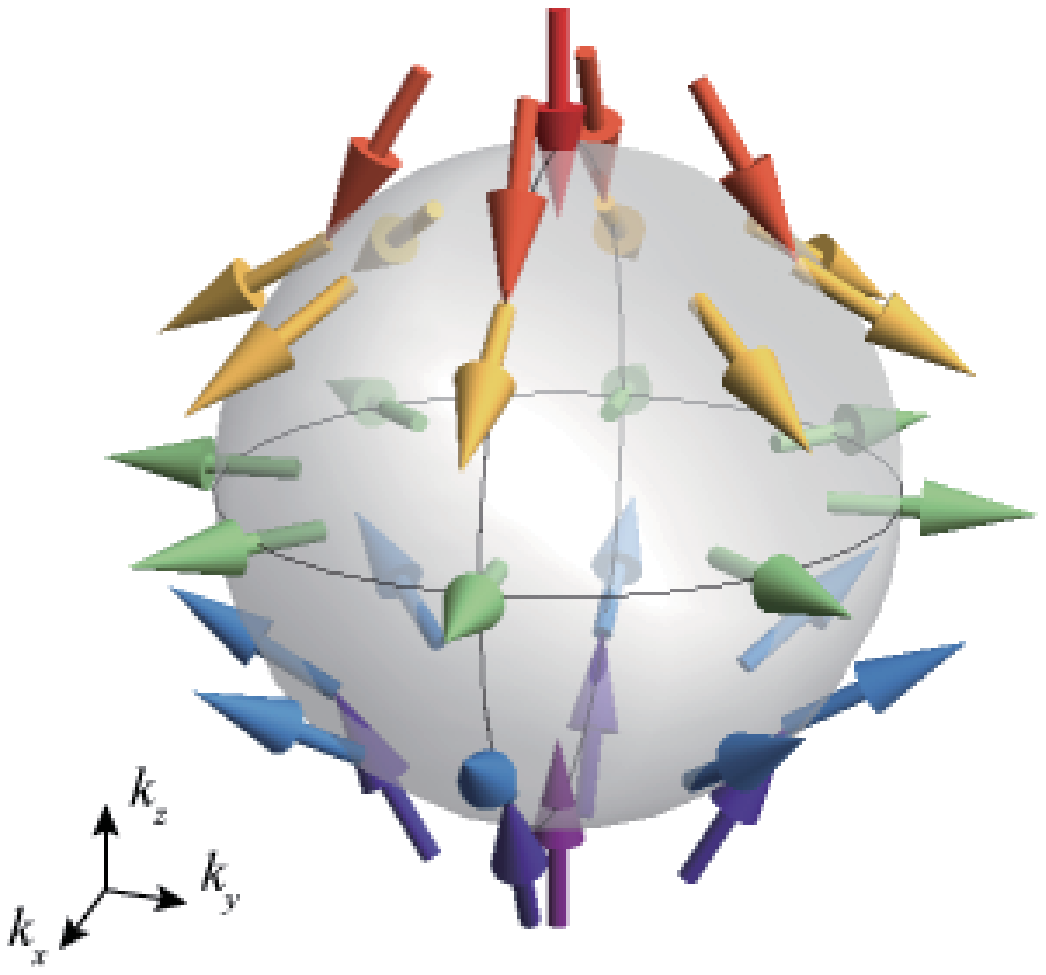}
\caption{(Color online) Vector field plot of the valence-band component of $\tilde {\bm d}{(\bm k)}$ given by eq. \ref{del2d} for $\Delta_2$ [(2,2)-component].
}
\label{d222}
\end{figure}

\begin{figure}
\centering
\includegraphics[scale=0.55]{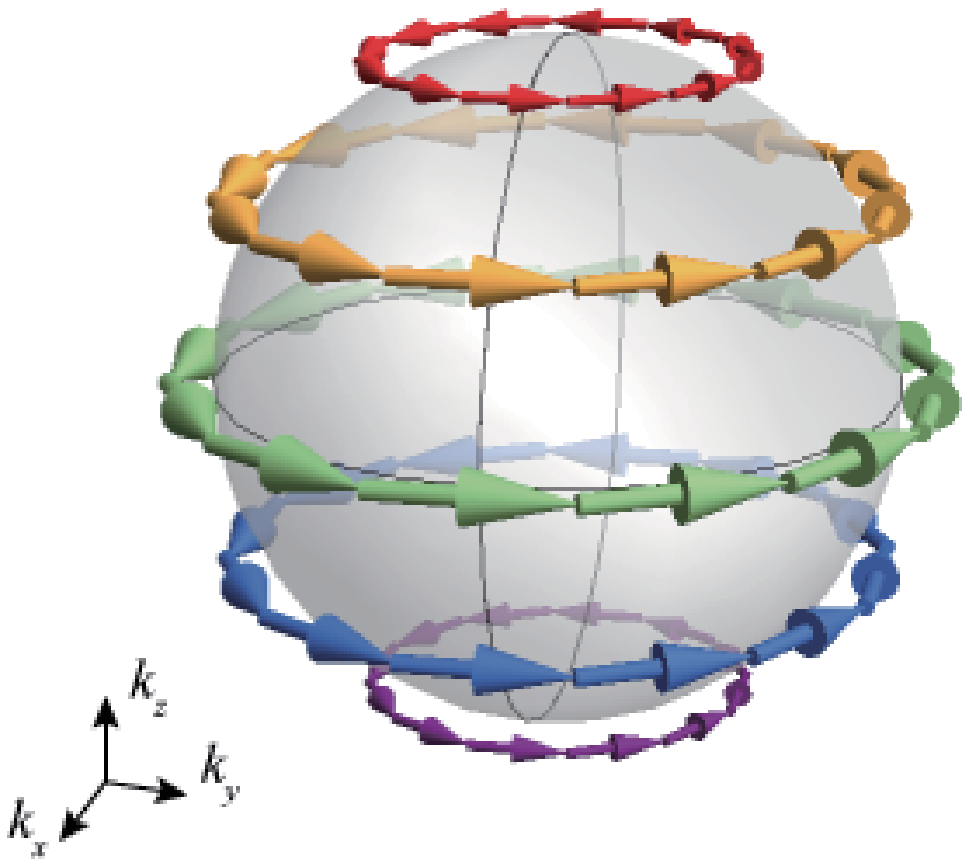}
\caption{
(Color online) Vector field plot of the conduction-band component of $\tilde {\bm d}{(\bm k)}$ given by eq. \ref{del3d} for $\Delta_3$ [(1,1)-component].
}
\label{d311}
\end{figure}

\begin{figure}
\centering
\includegraphics[scale=0.55]{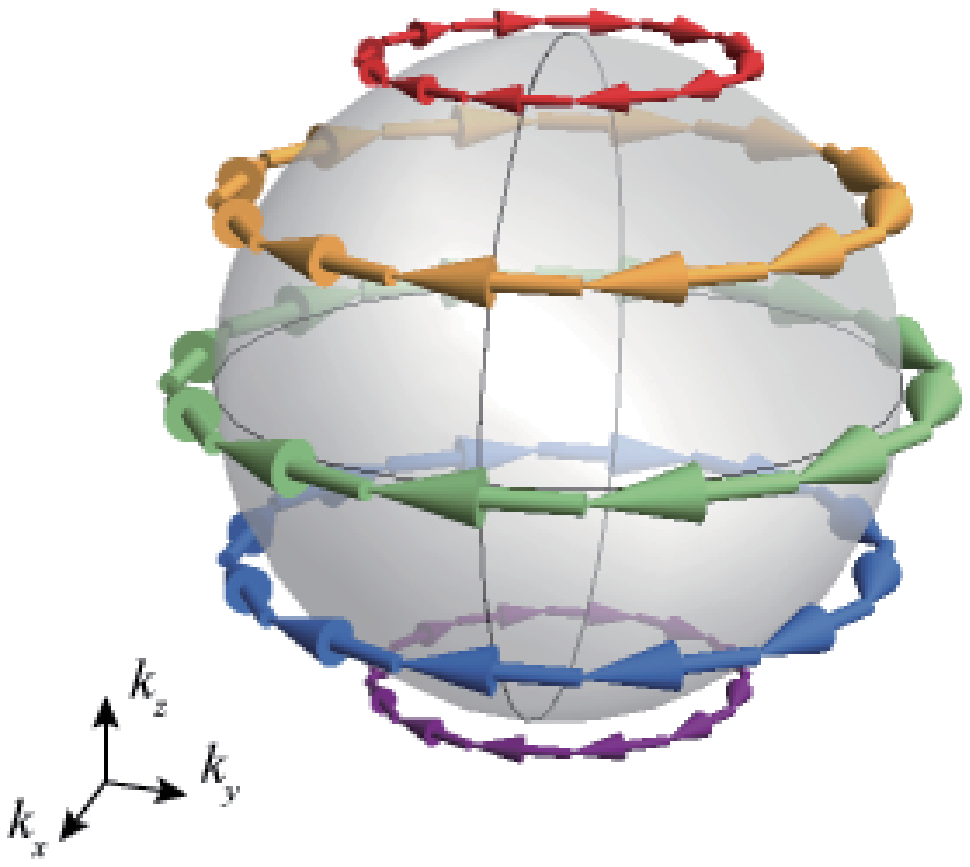}
\caption{
(Color online) Vector field plot of the valence-band component of $\tilde {\bm d}{(\bm k)}$ given by eq. (\ref{del3d}) for $\Delta_3$ [(2,2)-component].
}
\label{d322}
\end{figure}

\begin{figure}
\centering
\includegraphics[scale=0.55]{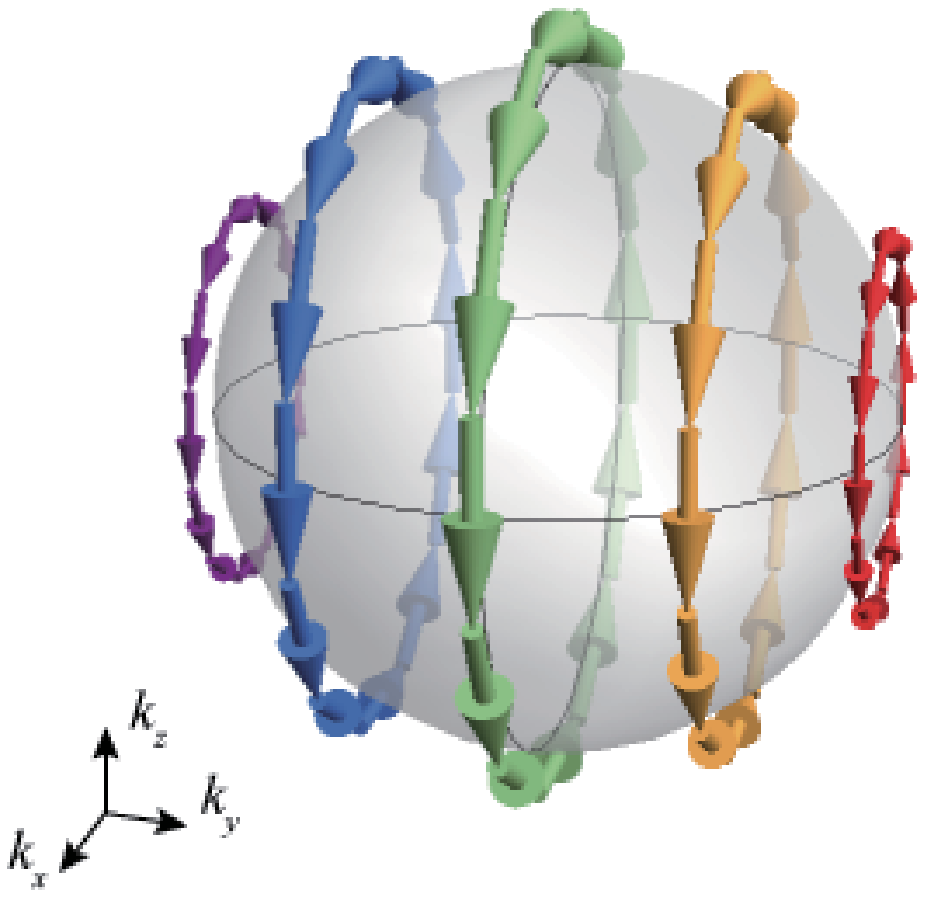}
\caption{
(Color online) Vector field plot of the conduction-band component of $\tilde {\bm d}{(\bm k)}$ given by eq. \ref{del4d} for $\Delta_4$ [(1,1)-component].
}
\label{d411}
\end{figure}

\begin{figure}
\centering
\includegraphics[scale=0.55]{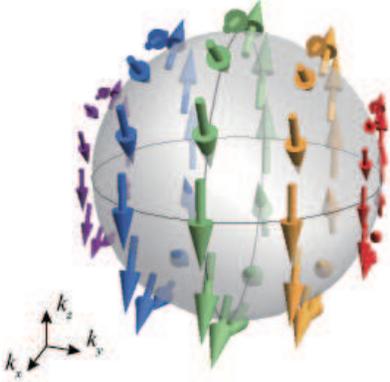}
\caption{
(Color online) Vector field plot of the valence-band component of $\tilde {\bm d}{(\bm k)}$ given by eq. \ref{del4d} for $\Delta_4$ [(2,2)-component].
}
\label{d422}
\end{figure}

In the band basis, spin-singlet and spin-triplet pair potentials can mix with each other because of the spin-orbit interaction.
Owing to this, 
it is rather difficult to understand
the temperature dependence of spin susceptibility.
In order to clarify the spin structures of pair potentials, it is necessary to introduce the band basis where the normal part of the Hamiltonian is diagonalized.
First, we diagonalize the Hamiltonian $H_0(\bm k)$ of the normal state as
\begin{align}
 H_0 (\bm k) \bm u_\gamma(\bm k) = \epsilon_\gamma(\bm k) \bm u_\gamma(\bm k),
\end{align}
with band index $\gamma$.
By using the unitary matrix $U(\bm k)$ given by $U(\bm k) = (\bm u_1(\bm k), \bm u_2(\bm k), \cdots)$, the pair potential $\hat \Delta = \sum_\mu d_\mu s_\mu$ is transformed as 
\begin{align}
\sum_\mu d_\mu s_\mu \to U^\dag(\bm k) \sum_\mu d_\mu s_\mu U(\bm k) = \sum_\mu \tilde d_\mu(\bm k) s_\mu.
\end{align}
$\tilde{d}_0(\bm k)$ 
and 
$\tilde {\bm{d}}(\bm k)$
denote the spin-singlet component of the pair potential and the $d$-vector of the spin-triplet component of pair potentials in the band basis, respectively.
Note that $s_\mu$ is not changed by this unitary transformation
in inversion-symmetric systems.
The corresponding Hamiltonian is expressed as 
\begin{align}
 U^\dag(\bm k) H(\bm k) U(\bm k)
 &= \mathrm{diag} (\epsilon_1(\bm k)-\mu, \epsilon_2(\bm k)-\mu, \cdots) \tau_z
\nonumber \\ & \qquad
+ \sum_\mu \tilde d_\mu(\bm k) s_\mu \tau_x.
\end{align}
In the following, we give the $d$-vectors in the band basis for the
lowest order of ${\bm k}$. 
The detailed derivation of $\tilde d_\mu (\bm k)$ is shown in Appendix. 
In the case of $\Delta_1$, $\tilde d_\mu (\bm k)$ is the same as that in the orbital basis: $d_0(\bm k) = \Delta, \bm d(\bm k) = \bm 0$.
For the other cases, we have
\\
$\Delta_2:$
\begin{align}
 \tilde d_0(\bm k)
 &= 0,\label{del2d0}
 \\
 \tilde{\bm d}(\bm k)
 &=\Delta 
\left(\frac{v k_x}{m_0}, \frac{v k_y}{m_0}, \frac{v_z k_z}{|m_0|} \tilde \sigma_z - \mathrm{sgn}(m_0) \tilde \sigma_y
\right),
\label{del2d}
\end{align}
$\Delta_3:$
\begin{align}
 \tilde d_0(\bm k)
 &= \Delta \tilde \sigma_x,\label{del3d0}
 \\
 \tilde{\bm d}(\bm k)
 &= \Delta \tilde \sigma_z 
\left(-\frac{v k_y}{|m_0|}, \frac{v k_x}{|m_0|}, 0
\right),
\label{del3d}
\end{align}
$\Delta_4:$
\begin{align}
 \tilde d_0(\bm k)
 &= \Delta \frac{v k_y}{m_0}\frac{v_z k_z}{m_0}\tilde\sigma_x,
 \label{del4d0}
 \\
 \tilde{\bm d}(\bm k)
 &= \Delta 
\left(\frac{v_z k_z}{|m_0|} \tilde \sigma_z - \tilde \sigma_y, 0, 
-\frac{v k_x}{m_0}
\right),
\label{del4d}
\end{align}
Here, 
$\tilde \sigma_i$ is the Pauli matrix denoting the band index, i.e., $\tilde \sigma_z=1$ for the conduction band and $\tilde \sigma_z = -1$ for the valence band.
To illustrate $\tilde{d}({\bm k})$ given by eq. (\ref{del2d}) in the
conduction (valence) band,
we plot 
the $(1,1)$-component
[$(2,2)$-component] of the $\tilde{d}$ vector in Fig. \ref{d211}
(Fig. \ref{d222}).
Those given by eqs. (\ref{del3d}) and  (\ref{del4d}) are also shown in Figs. \ref{d311} and \ref{d322} and in Figs. \ref{d411} and \ref{d422}, respectively.
$\tilde d_\mu(\bm k)$ is useful for understanding the temperature dependence of $\chi_i$, as we will see in the following.\\
\subsection{Isotropic full-gap $\Delta_1$: Van Vleck susceptibility}
\begin{figure*}
\centering
\includegraphics{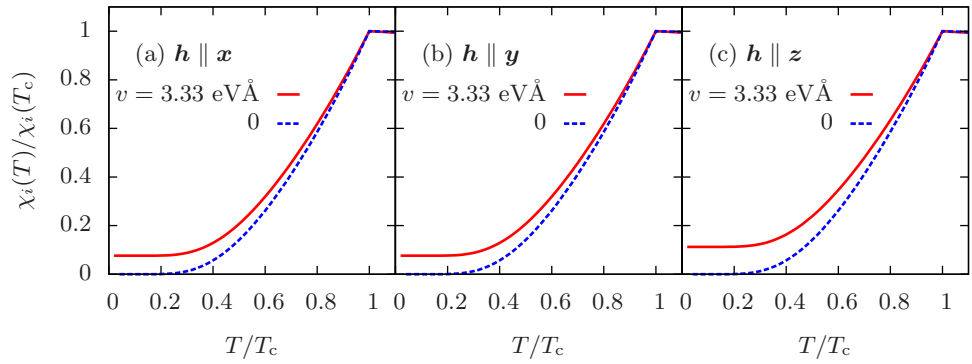}
\caption{(Color online) Tesmperature dependences of spin susceptibilities $\chi_x$ (a), $\chi_y$ (b), and $\chi_z$ (c) of STI with $\Delta_1$ in the presence ($v=3.33$ eV \AA, solid line) and absence ($v=0$, dashed line) of the spin-orbit interaction.
The value of $\chi_i$ is normalized by that in the normal state, which is given by $\chi_x(T_{\rm c}) = \chi_y({T_{\rm c}}) = 0.309 \chi_z(T_{\rm c})$ for $v=3.33$ eV \AA\ and $\chi_x (T_{\rm c}) = \chi_y (T_{\rm c}) = 0.210 \chi_z(T_{\rm c})$ for $v=0$.
}
\label{chi1}
\end{figure*}

Figure \ref{chi1} shows the temperature dependence of $\chi_i$  with $\Delta_1$ for $v=3.33$ eV \AA\ and $v=0$, where $v$ corresponds to the strength of the spin-orbit interaction.
An STI with $\Delta_1$ is a full-gap superconductor, therefore
resulting in $\chi_x$, $\chi_y$, and $\chi_z$ decreasing exponentially with decreasing $T$ for $T < T_{\rm c}$ for both $v=0$ and $v=3.33$ eV \AA.
In the case of $v=0$,
all the $\chi_i$ vanish at $T=0$ as shown by the dashed lines in Fig. \ref{chi1}.
On the other hand, in the presence of the spin-orbit interaction, all the $\chi_i$ remain at a finite value at $T=0$ (solid line in Fig. \ref{chi1}) owing to the Van Vleck susceptibility,\cite{Maruyama} which is allowed in a multi band system with the spin-orbit interaction.
Actually, $\chi_z$ at $T=0$ is proportional to $v^2$ (see Appendix \ref{vanvleck}).
Note that
the value of $\chi_z$ is larger than those of $\chi_x$ and $\chi_y$
 in the normal state 
because of the anisotropy of the energy band.

\subsection{Anisotropic full-gap $\Delta_2$: Rotation of $d$-vector}
The $d$-vector in an STI with $\Delta_2$ is parallel to the $z$-axis in the orbital basis.
In the absence of the spin-orbit interaction, the $d$-vector for $\Delta_2$ in the band basis is also parallel to the $z$-axis as shown by eq. (\ref{del2d}).
Consequently, only $\chi_z$ decreases with decreasing $T$ and vanishes at $T=0$, and $\chi_x$ and $\chi_y$ are independent of $T$, as shown in Fig. \ref{chi2}. 
At low temperatures, $\chi_z$ is proportional to $T$ since an STI with $\Delta_2$ has a line node on the equator for $v=0$, as discussed in \S. \ref{model}.
In the presence of the spin-orbit interaction,
$\chi_z$ decreases exponentially with decreasing $T$ for $T<T_{\rm c}$, as denoted by the solid line in Fig. \ref{chi2}(c).
In addition,
$\chi_x$ and $\chi_y$ slightly decrease 
[solid lines in Figs. \ref{chi2}(a) and 13(b)] with decreasing $T$ since 
the $d$-vector is rotated so that the $d_x$- and $d_y$-components are induced in the band basis. $\chi_z$ at $T=0$ takes a finite value for the following two reasons.
First, $\tilde {\bm d}(\bm k)$ is not parallel to the $z$-axis.
Second, the Van Vleck susceptibility arises, as in the case of $\Delta_1$.

\subsection{Point node on poles $\Delta_3$: Induced spin-triplet pair potential}
\begin{figure*}
\centering
\includegraphics{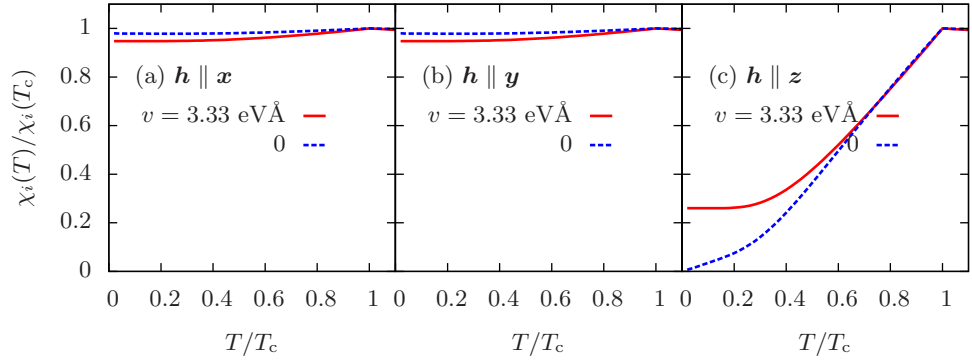}
\caption{(Color online) Temperature dependences of spin susceptibilities of STI with $\Delta_2$.
}
\label{chi2}
\end{figure*}

\begin{figure*}
\centering
\includegraphics{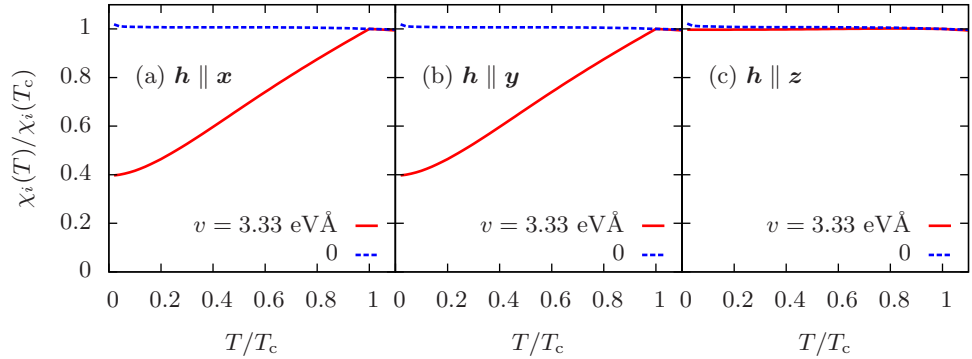}
\caption{(Color online) Temperature dependences of spin susceptibilities of STI with $\Delta_3$.}
\label{chi3}
\end{figure*}

For $v$=0, all the $\chi_i$ of an STI with $\Delta_3$ are independent of $T$, as denoted by the dashed line in Fig. \ref{chi3}, since the energy spectrum is gapless (\S \ref{sped3}).
On the other hand, for $v=3.33$ eV\AA, $\chi_x$ and $\chi_y$ decrease with decreasing $T$ to $\chi_i(T=0)/\chi_i(T_c) \sim 0.4$ at $T=0$ [solid lines in Fig. \ref{chi3}(a)(b)], while $\chi_z$ is independent of $T$ [solid line in Fig. \ref{chi2}(c)].
This behavior can be understood from the induced spin-triplet component
$\tilde {\bm d}(\bm k)$ in eq. (\ref{del3d}) owing to the spin-orbit interaction.
The induced $d$-vector $\tilde {\bm d}(\bm k)$ is parallel to the $xy$-plane, as shown in Figs. \ref{d311} and \ref{d322}; consequently, $\chi_z$ becomes independent of $T$.
Moreover, $\chi_x$ and $\chi_y$ take finite values at $T=0$ owing to the Van Vleck susceptibility. 
A similar result  has been obtained for a bilayer system.\cite{Maruyama}

\subsection{Point nodes on equator $\Delta_4$: Rotation of $d$-vector and induced spin-singlet pair potential}

\begin{figure*}
\centering
\includegraphics{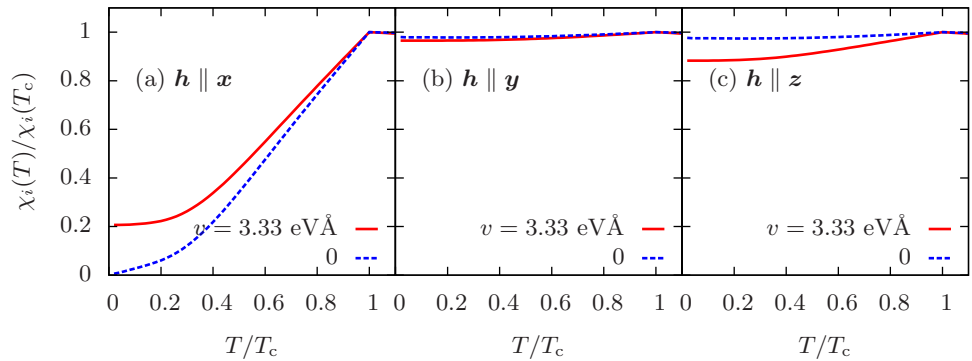}
\caption{(Color online) Temperature dependences of spin susceptibilities of STI with $\Delta_4$.}
\label{chi4}
\end{figure*}

For $v=0$, because $\tilde {\bm d}(\bm k) \parallel \bm x$ in the band basis, only $\chi_x$ decreases with decreasing $T$ for $T<T_{\rm c}$ and vanishes at $T=0$ [dashed line in Fig. \ref{chi4}(a)].
At low temperatures, $\chi_x$ is proportional to $T$ since the energy spectrum has a line node on the equator (see \S. \ref{model}). 
For $v=3.33$ eV\AA,
$\chi_x$ decreases with decreasing $T$ and is proportional to $T^2$ at low temperatures, except for the residual value at $T=0$.
This residual spin susceptibility originates from the rotated $d$-vector and the Van Vleck susceptibility due to the spin-orbit interaction.
$\chi_z$ slightly decreases with decreasing $T$ for $T<T_{\rm c}$
since $\tilde d_z(\bm k)$ is present.
On the other hand,
$\tilde d_y(\bm k)$ vanishes up to the first order of $\bm k$
[eq. (\ref{del4d}) and Figs. \ref{d411} and \ref{d422}], and thus
$\chi_y$ is almost independent of $T$ [solid line in Fig. \ref{chi4}(b)].

\section{Discussion and Summary}
\label{summary}

In this paper, 
we have calculated the temperature dependence of the specific heat and the spin susceptibility.
The temperature dependences of the specific heat are similar among three
of the four possible pair potentials.
On the other hand, wide variations of the temperature dependence appear in the spin susceptibility depending on the direction of the applied magnetic field.
These results are summarized in Table \ref{table3}.

Finally, we compare the obtained results and the experimental ones.
From the temperature dependence of the specific heat, 
$\Delta_1$, $\Delta_2$, and $\Delta_4$ are almost consistent with the experimental result.
On the other hand, in a recent tunneling spectroscopy investigation of the $(111)$ surface of Cu$_x$Bi$_2$Se$_3$, a pronounced ZBCP was obserbed.\cite{Sasaki}
From the theoretical calculation, the ZBCP due to ABSs is generated on
the (111) surface only for the case with $\Delta_2$ and
$\Delta_4$.\cite{Sasaki,Yamakage1}
On the basis of this background, the promising pair potentials are $\Delta_2$ and
$\Delta_4$.
In the light of the obtained spin susceptibility in this paper,
we conclude that it is possible to distinguish between $\Delta_2$ and
$\Delta_4$ by measuring the temperature dependence of the in-plane and
out-of-plane Knight shifts.

\begin{center}
\begin{table}
\begin{tabular}{l|c|c|ccc} \hline\hline

$\ \ \ $pairing & specific heat & Andreev & \multicolumn{3}{c}{spin susceptibility} \\
$\ \ $potential&$C_{s}$&bound state&$\ \ \ $$\chi_{x}$$\ \ \ $&$\chi_{y}$&$\chi_{z}$\\ \hline
$\Delta_{1} = \Delta$ & yes & no & $\swarrow$ & $\swarrow$ & $\swarrow$ \\  
$\Delta_{2} = \Delta\sigma_y s_z$ & yes & yes & $-$ & $-$ & $\swarrow$ \\ 
$\Delta_{3} = \Delta\sigma_z $ & no & no & $\swarrow$ & $\swarrow$ & $-$ \\ 
$\Delta_{4} = \Delta\sigma_y s_x$ & yes  & yes & $\swarrow$ & $-$ & $-$ \\ \hline\hline

\end{tabular}
\caption{
[First column] Possible pairing symmetry of an STI. 
[Second column] Comparison of line shape between our results and the experimental one.\cite{Kriener1}
[Third column] Presence or absence of ZBCP due to ABSs.\cite{Sasaki,Yamakage1}
[Fourth column] Temperature dependence of $\chi_x$, $\chi_y$, and $\chi_z$.
$\swarrow$ denotes a decrease in $\chi_i$ with decreasing temperature.
$-$ denotes that $\chi_i$ is almost independent of the temperature.}
{\label{table3}}
\end{table}
\end{center}

\section{Acknowledgements}

We are grateful to M. Kriener, K. Segawa, Z. Ren, S. Sasaki and Y. Ando for valuable discussions and
providing the experimental data. We gratefully acknowledge S. Onari
for valuable discussions.
This work was supported in part by Grants-in-Aid
for Scientific Research from MEXT of Japan ``Topological
Quantum Phenomenah(Grant Nos. 22103005, 20654030 and 22540383).
\appendix

\section{Effects of Spin-Orbit Interaction}

\subsection{Van Vleck susceptibility for $\Delta_1$}
\label{vanvleck}
Here, we derive the Van Vleck susceptibility, which gives a finite value of the spin susceptibility at $T=0$.
We focus on an STI with $\Delta_1$, based on the Hamiltonian in the continuum limit given by 
\begin{align}
 H(\bm k) &= [c(\bm k) + m(\bm k) \sigma_x + v_z k_z \sigma_y 
+ v k_{\parallel}
h_s(\bm k) \sigma_z 
]\tau_z 
+ \Delta \tau_x.
\end{align}
First, we diagonalize the spin part: $h_{s}(\bm k) = (\bm k \times \bm s)_z/k_{\parallel}$, 
where $k_{\parallel} = |\bm k_\parallel| = (k_x^2 + k_y^2)^{1/2}$.
The eigenvalue $\tilde s$ of $h_{s}$ is given by $\tilde s=\pm 1$.
The corresponding eigenvector is given by
\begin{align}
 |\tilde s \rangle = 
\frac{1}{\sqrt 2}
\begin{pmatrix}
   1
   \\
   \tilde s e^{i \varphi_{\bm k}}
 \end{pmatrix},
 \label{vector_helicity}
\end{align}
with 
$\sin \varphi_{\bm k} = k_x/k_{\parallel}$ and $\cos \varphi_{\bm k} = -k_y/k_{\parallel}$.

Next, we diagonalize the normal part: $H_{0 \tilde s}(\bm k) = m(\bm k) \sigma_x + v_z k_z \sigma_y + \tilde s v k_\parallel \sigma_z$.
The eigenvalue of $H_{0 \tilde s}(\bm k)$ is given by $\tilde \sigma \eta(\bm k) + c(\bm k)$ with
\begin{align}
 \eta(\bm k) = \sqrt{m^2(\bm k) + v_z^2 k_z^2 + v^2 k^2_{\parallel}},
\end{align}
where $\tilde \sigma = \pm 1$ is the band index.
The corresponding eigenvectors $| \tilde s \tilde \sigma \rangle$ of $H_{0 \tilde s}(\bm k)$ are given by
\begin{align}
 |\pm, \pm \rangle
 &= \begin{pmatrix}
   \cos p_{\bm k}/2
   \\
   \pm e^{i q_{\bm k}} \sin p_{\bm k}/2
 \end{pmatrix},\label{vector_band1}
\\
 |\pm, \mp \rangle
 &= \begin{pmatrix}
   \sin p_{\bm k}/2
   \\
   \mp e^{i q_{\bm k}}\cos p_{\bm k}/2
 \end{pmatrix},
 \label{vector_band2}
\end{align}
with 
\begin{align}
\cos p_{\bm k} &= vk_{\parallel}/\eta(\bm k),
\\
\sin p_{\bm k} &= \sqrt{m^2(\bm k) + v_z^2 k_z^2}/\eta(\bm k),
\\
\cos q_{\bm k} &= m(\bm k)/\sqrt{m^2(\bm k) + v_z^2 k_z^2},
\\
\sin q_{\bm k} &= v_z k_z/\sqrt{m^2(\bm k) + v_z^2 k_z^2}.
\end{align}
%
%
In the band basis, the original Hamiltonian is rewritten as
\begin{align}
 H_{\tilde s \tilde \sigma}(\bm k) = [ \tilde \sigma \eta(\bm k) +c(\bm k) ]\tau_z + \Delta \tau_x.
\end{align}
The energy eigenvalue of $H_{\tilde s \tilde \sigma}(\bm k)$ is given by $\tau E_{\tilde \sigma}(\bm k)$ with
\begin{align}
 E_{\tilde \sigma}(\bm k) =  \sqrt{[\tilde\sigma \eta(\bm k) + c(\bm k) ]^2 + \Delta^2},
\end{align}
and $\tau = \pm 1$.
The corresponding eigenvectors $| \tilde s \tilde \sigma \tau \rangle$ are given by
\begin{align}
 | \tilde s, \tilde \sigma, + \rangle 
 &= \begin{pmatrix}
   \cos P_{\bm k \tilde \sigma}/2
   \\
   \sin P_{\bm k \tilde \sigma}/2
 \end{pmatrix},
 \\
 | \tilde s, \tilde \sigma, - \rangle 
 &= \begin{pmatrix}
   \sin P_{\bm k \tilde \sigma}/2
   \\
   -\cos P_{\bm k \tilde \sigma}/2
 \end{pmatrix},
\end{align}
with $\cos P_{\bm k \tilde \sigma} = [\tilde \sigma \eta(\bm k) + c(\bm k)] / E_{\tilde \sigma}(\bm k)$.

For a full-gap system, the spin susceptibility at $T=0$ is given by
\begin{align}
 \chi_i &= \frac{2\mu_{\rm B}^2 }{N}
 \sum_{\bm k \tilde s \tilde s'\tilde \sigma \tilde \sigma' }
 \frac{
\left
|\langle \tilde s | s_i | \tilde s' \rangle
\langle \tilde s, \tilde \sigma | \tilde s', \tilde \sigma'
\rangle
\langle
\tilde s, \tilde \sigma, + | \tilde s', \tilde \sigma', -
\rangle
\right|^2}
{ E_{\tilde \sigma}(\bm k) +  E_{\tilde \sigma'}(\bm k)}.
\end{align}
For simplicity, we assume that all the $g$-factors are equal to two, and we concentrate on $\chi_z$.
The matrix elements in the above expression are estimated as follows.
\begin{align}
 \langle \tilde s | s_z | \tilde s' \rangle &= 1 - \delta_{\tilde s \tilde s'},
 \label{s-sz-s}
 \\
 \langle \tilde s, \pm | -\tilde s, \mp \rangle
 &= \pm \cos p_{\bm k},
 \\
 \langle \tilde s, \tilde \sigma, + | -\tilde s, \tilde \sigma', - \rangle
 &= \sin \frac{P_{\bm k \tilde \sigma'} - P_{\bm k \tilde \sigma}}{2}.
\end{align}
From the above equations, only the Van Vleck term, which originates from the off-diagonal terms of $\tilde s \ne \tilde s'$, $\tilde \sigma \ne \tilde \sigma'$, and $\tau \ne \tau'$, can be nonzero and is given as
\begin{align}
 \chi_z &= \frac{8\mu_{\rm B}^2}{N}
 \sum_{\bm k}
 \frac{1}{E_{+}(\bm k) + E_{-}(\bm k)}
 \frac{v^2 k_{\parallel}^2}{\eta^2(\bm k)}
 \sin^2 \frac{P_{\bm k +} - P_{\bm k -}}{2}.
\end{align}
One can verify that $\chi_z \to 0$ as $v \to 0$ from the above expression.
The spin susceptibility at $T=0$ in an STI with $\Delta_1$ stems from the Van Vleck component due to the spin-orbit coupling $v$.

\subsection{Rotation of $d$-vector for $\Delta_2$}
\label{rotation}

Here, we derive the $d$-vector for $\Delta_2$ in the band basis.
The following relation is useful:
\begin{align}
 \langle \tilde s, \tilde \sigma | \sigma_y | \tilde s', \tilde \sigma' \rangle
 &= 
\bigl[
\delta_{\tilde s \tilde s'}
 \left(
   \tilde \sigma_z \sin p_{\bm k} \sin q_{\bm k} - \tilde \sigma_y \cos q_{\bm k}
 \right)
 \nonumber\\ & \quad
 - (\tilde s_z)_{\tilde s \tilde s'}
 \tilde \sigma_x \cos p_{\bm k} \sin q_{\bm k}
 \nonumber\\ & \quad
 + (\tilde s_x)_{\tilde s \tilde s'}
 \left(
   \tilde \sigma_z \sin q_{\bm k} - \tilde \sigma_y \sin p_{\bm k} \cos q_{\bm k}
 \right)
 \nonumber\\ & \quad
 + (\tilde s_y)_{\tilde s \tilde s'} \sigma_0 \cos p_{\bm k} \cos q_{\bm k}
\bigr]_{\tilde \sigma \tilde \sigma'}.
\label{sigmay}
\end{align}
The above expression is derived using eqs. (\ref{vector_band1}) and (\ref{vector_band2}).
The pair potential $\Delta_2$ is represented in the band basis as
\begin{align}
\Delta  \sigma_y  s_z
&=
 \Delta(\sin q_{\bm k} \tilde\sigma_z - \cos q_{\bm k} \sin p_{\bm k} \tilde \sigma_y)\tilde s_x
+
 \Delta\cos q_{\bm k} \frac{vk_\parallel}{\eta(\bm k)} \tilde s_y,
\end{align}
where 
$\tilde s_i$ is the Pauli matrix in the spin-helicity space.
Here, the relation between $\bm s$ and $\tilde {\bm  s}$
is as follows:
\begin{align}
 \langle \tilde s | s_z | \tilde s' \rangle 
&= 1-\delta_{\tilde s \tilde s'} = (\tilde s_x)_{\tilde s \tilde s'}
 \\
 \langle \tilde s | s_x | \tilde s' \rangle 
&= (\tilde s_y)_{\tilde s \tilde s'} \sin \varphi_{\bm k} + (\tilde s_z)_{\tilde s \tilde s'} \cos \varphi_{\bm k},
\label{sx}
\\
 \langle \tilde s | s_y | \tilde s' \rangle 
&= (\tilde s_z)_{\tilde s \tilde s'} \sin \varphi_{\bm k} - (\tilde s_y)_{\tilde s \tilde s'} \cos \varphi_{\bm k},
\end{align}
or equivalently,
\begin{align}
 \tilde s_x &= s_z,\label{tilde_sx}
\\
\tilde s_y &= s_x \sin \varphi_{\bm k}  - s_y \cos \varphi_{\bm k}
= \frac{\bm k_{\parallel} \cdot \bm s}{k_\parallel},\label{tilde_sy}
\\
\tilde s_z &= s_x \cos \varphi_{\bm k}  + s_y \sin \varphi_{\bm k}
= \frac{(\bm k \times \bm s)_z}{k_\parallel}.
\label{tilde_sz}
\end{align}
Consequently, the $d$-vector for $\Delta_2$ in the band basis is obtained as
\begin{align}
 \tilde d_x(\bm k)
 &= \Delta \cos q_{\bm k} \frac{vk_x}{\eta(\bm k)},
 \\
 \tilde d_y(\bm k)
 &= \Delta \cos q_{\bm k} \frac{vk_y}{\eta(\bm k)},
 \\
 \tilde d_z(\bm k)
 &= \Delta (\sin q_{\bm k} \tilde \sigma_z - \cos q_{\bm k} \sin p_{\bm k} \tilde \sigma_y).
\end{align}
Note that the spin in the above expression is represented in the original spin space ($\bm s$) not in the  spin-helicity space ($\tilde {\bm s}$).
For an STI with $\Delta_2$,
the spin-orbit interaction has the role of rotating the $d$-vector in the band basis.
In the case of $v=0$,
because $\tilde {\bm d}(\bm k) \parallel  \bm z$,
only $\chi_z$ decreases with decreasing $T$ for $T<T_{\rm c}$.
In the case of $v \ne 0$, $\tilde d_x(\bm k)$ and $\tilde d_y(\bm k)$ (proportional to $v	$) are present, and $\chi_x$ and $\chi_y$ also decrease slightly with decreasing $T$ for $T<T_{\rm c}$.

\subsection{Induced spin-triplet pair for $\Delta_3$}

In the following, we show that a spin-triplet pair is induced for $\Delta_3$ in the band basis.
As in Appendix \ref{rotation},
we derive $\tilde {\bm d}(\bm k)$ for $\Delta_3 = \Delta \sigma_z$.
Using eqs. (\ref{vector_band1}) and (\ref{vector_band2}),
the matrix elements of $\sigma_z$ are obtained as
\begin{align}
 \langle \tilde s, \tilde \sigma | \sigma_z | \tilde s, \tilde \sigma' \rangle
 = 
\left(
\tilde s \tilde \sigma_z \cos p_{\bm k} + \tilde \sigma_x \sin p_{\bm k}
\right)_{\tilde \sigma \tilde \sigma'}.
\end{align}
Therefore, $\sigma_z$ is expressed in the band basis as
\begin{align}
 \sigma_z 
&=
\sin p_{\bm k} \tilde \sigma_x +  \frac{v (k_x s_y - k_y s_x)}{\eta(\bm k)} \tilde \sigma_z.
\end{align}
This is derived with the help of eq. (\ref{tilde_sz}).
As a result, $\tilde d_0(\bm k)$ and $\tilde {\bm d}(\bm k)$ are given by
\begin{align}
 \tilde d_0(\bm k) 
&=
\Delta \sin p_{\bm k} \tilde \sigma_x,
\\
 \tilde {\bm d}(\bm k) 
&= \frac{v \Delta}{\eta(\bm k)} \tilde \sigma_z (-k_y,k_x,0),
\end{align}
which implies that
a spin-triplet pair is induced in the band basis.
Note that $\tilde {\bm d}(\bm k) \perp \bm z$.
This is the reason why almost only $\chi_z$ in an STI with $\Delta_3$
decreases with decreasing temperature.

\subsection{Induced spin-singlet pair and rotation of $d$-vector for $\Delta_4$}

In this subsection, we derive the $d$-vector for $\Delta_4=\Delta\sigma_y s_x$,
and show that a spin-singlet pair is induced and that the $d$-vector is rotated in the band basis. 
From the matrix elements of $\sigma_y$ [eq. (\ref{sigmay})] and $s_x$ [eq. (\ref{sx})],the pair potential $\Delta_4$ in the band basis is represented as follows:
\begin{align}
\Delta \sigma_y s_x 
&=
\Delta \frac{v k_y}{\eta(\bm k)} \sin q_{\bm k} \tilde \sigma_x 
- \Delta \frac{vk_x}{\eta(\bm k)} \cos q_{\bm k}\tilde s_x 
\nonumber\\ & \quad
+
\Delta
\frac{k_x}{k_\parallel}
( \sin q_{\bm k} \tilde \sigma_z - \sin p_{\bm k} \cos q_{\bm k}  \tilde \sigma_
y)\tilde s_y 
\nonumber\\ & \quad
-\Delta 
\frac{k_y}{k_\parallel}
(
\sin p_{\bm k}
\sin q_{\bm k}  \tilde \sigma_z -  \cos q_{\bm k} \tilde \sigma_y) \tilde s_z.
\end{align}
Using eqs. (\ref{tilde_sx}) - (\ref{tilde_sz}), the $d$-vector for $\Delta_4$ in the band basis is obtained as
\begin{align}
 \tilde d_0(\bm k)
 &= 
\Delta \frac{v k_y}{\eta(\bm k)}\sin q_{\bm k} \tilde \sigma_x,
 \\
 \tilde d_x(\bm k)
 &= \Delta 
\Bigg[
\Biggl(
\frac{k_y^2}{k^2_\parallel} \frac{v_z k_z}{\eta(\bm k)} + \frac{k_x^2}{k^2_\parallel} \sin q_{\bm k}
\Biggr) \tilde \sigma_z \nonumber
 \\
 &\hspace{1em} 
- 
\left(\frac{k_x^2}{k_\parallel^2} 
\frac{m(\bm k)}{\eta(\bm k)}  + 
\frac{k_y^2}{k_\parallel^2}
 \cos q_{\bm k}
\right)\tilde \sigma_y
\Biggr],
 \\
 \tilde d_y(\bm k)
 &= \Delta \frac{k_xk_y}{k_\parallel^2} 
\biggl[
\left(\sin q_{\bm k} - \frac{v_z k_z}{\eta(\bm k)} 
\right) \tilde \sigma_z 
+ 
\left(\cos q_{\bm k} - \frac{m(\bm k)}{\eta(\bm k)} 
\right) \tilde \sigma_y
\biggr], 
 \\
 \tilde d_z(\bm k)
 &= - \Delta \frac{vk_x}{\eta(\bm k)} \cos q_{\bm k}.
\end{align}
Therefore, because of the spin-orbit interaction, 
a spin singlet pair $\tilde d_0(\bm k)$ is induced in the band basis,
and 
the $d$-vector is rotated so that $\tilde d_y(\bm k)$ and $\tilde
d_z(\bm k)$ become nonzero.

\bibliography{refer}

\begin{thebibliography}{10}

\bibitem{Hasan}
M.~Z. Hasan and C.~L. Kane: Rev. Mod. Phys. {\bfseries 82} (2010) 3045.

\bibitem{XLQi}
X.-L. Qi and S.-C. Zhang: Rev. Mod. Phys. {\bfseries 83} (2011) 1057.

\bibitem{Tanaka1}
Y.~Tanaka, N.~Nagaosa, and M.~Sato: J. Phys. Soc. Jpn. {\bfseries 81} (2012)
  011013.

\bibitem{Schnyder}
A.~P. Schnyder, S.~Ryu, A.~Furusaki, and A.~W.~W. Ludwig: Phys. Rev. B
  {\bfseries 78} (2008) 195125.

\bibitem{Sato09}
M.~Sato: Phys. Rev. B {\bfseries 79} (2009) 214526.

\bibitem{Sato10}
M.~Sato: Phys. Rev. B {\bfseries 81} (2010) 220504(R).

\bibitem{Wilczek}
F.~Wilczek: Nat. Phys. {\bfseries 5} (2009) 614.

\bibitem{Maeno}
Y.~Maeno, S.~Kittaka, T.~Nomura, S.~Yonezawa, and K.~Ishida: J. Phys. Soc. Jpn
  {\bfseries 81} (2012) 011009.

\bibitem{Mackenzie}
A.~P. Mackenzie and Y.~Maeno: Rev. Mod. Phys. {\bfseries 75} (2003) 657.

\bibitem{Furusaki}
A.~Furusaki, M.~Matsumoto, and M.~Sigrist: Phys. Rev. B {\bfseries 64} (2001)
  054514.

\bibitem{Stone}
M.~Stone and R.~Roy: Phys. Rev. B {\bfseries 69} (2004) 184511.

\bibitem{Kashiwaya}
S.~Kashiwaya, H.~Kashiwaya, H.~Kambara, T.~Furuta, H.~Yaguchi, Y.~Tanaka, and
  Y.~Maeno: Phys. Rev. Lett. {\bfseries 107} (2011) 077003.

\bibitem{Tanaka}
Y.~Tanaka, T.~Yokoyama, A.~V. Balatsky, and N.~Nagaosa: Phys. Rev. B {\bfseries
  79} (2009) 060505.

\bibitem{Sato1}
M.~Sato and S.~Fujimoto: Phys. Rev. B {\bfseries 79} (2009) 094504.

\bibitem{Sato2}
M.~Sato, Y.~Takahashi, and S.~Fujimoto: Phys. Rev. Lett. {\bfseries 103} (2009)
  020401.

\bibitem{Sato3}
M.~Sato, Y.~Takahashi, and S.~Fujimoto: Phys. Rev. B {\bfseries 82} (2010)
  134521.

\bibitem{Sato4}
M.~Sato and S.~Fujimoto: Phys. Rev. Lett. {\bfseries 105} (2010) 217001.

\bibitem{Sau}
J.~D. Sau, R.~M. Lutchyn, S.~Tewari, and S.~Das~Sarma: Phys. Rev. Lett.
  {\bfseries 104} (2010) 040502.

\bibitem{Alicea1}
J.~Alicea: Phys. Rev. B {\bfseries 81} (2010) 125318.

\bibitem{Lutchyn1}
R.~M. Lutchyn, J.~D. Sau, and S.~Das~Sarma: Phys. Rev. Lett. {\bfseries 105}
  (2010) 077001.

\bibitem{Oreg}
Y.~Oreg, G.~Refael, and F.~von Oppen: Phys. Rev. Lett. {\bfseries 105} (2010)
  177002.

\bibitem{Lutchyn2}
R.~M. Lutchyn, T.~D. Stanescu, and S.~Das~Sarma: Phys. Rev. Lett. {\bfseries
  106} (2011) 127001.

\bibitem{Alicea2}
J.~Alicea, Y.~Oreg, G.~Rafael, F.~von Oppen, and M.~F. Fisher: Nat. Phys.
  {\bfseries 7} (2011) 412.

\bibitem{FuKane1}
L.~Fu and C.~L. Kane: Phys. Rev. Lett. {\bfseries 100} (2008) 096407.

\bibitem{FuKane2}
L.~Fu and C.~L. Kane: Phys. Rev. Lett. {\bfseries 102} (2009) 216403.

\bibitem{Akhmerov}
A.~R. Akhmerov, J.~Nilsson, and C.~W.~J. Beenakker: Phys. Rev. Lett. {\bfseries
  102} (2009) 216404.

\bibitem{Law}
K.~T. Law, P.~A. Lee, and T.~K. Ng: Phys. Rev. Lett. {\bfseries 103} (2009)
  237001.

\bibitem{Tanaka2}
Y.~Tanaka, T.~Yokoyama, and N.~Nagaosa: Phys. Rev. Lett. {\bfseries 103} (2009)
  107002.

\bibitem{Linder}
J.~Linder, Y.~Tanaka, T.~Yokoyama, A.~Sudb\o{}, and N.~Nagaosa: Phys. Rev.
  Lett. {\bfseries 104} (2010) 067001.

\bibitem{Yamakage1}
A.~Yamakage, K.~Yada, M.~Sato, and Y.~Tanaka: Phys. Rev. B {\bfseries 85}
  (2012) 180509.

\bibitem{Yamakage2}
A.~Yamakage, Y.~Tanaka, and N.~Nagaosa: Phys. Rev. Lett. {\bfseries 108} (2012)
  087003.

\bibitem{Beenackker}
C.~W.~J. {Beenakker}: arXiv:1112.1950 .

\bibitem{Hor}
Y.~S. Hor, A.~J. Williams, J.~G. Checkelsky, P.~Roushan, J.~Seo, Q.~Xu, H.~W.
  Zandbergen, A.~Yazdani, N.~P. Ong, and R.~J. Cava: Phys. Rev. Lett.
  {\bfseries 104} (2010) 057001.

\bibitem{Sasaki}
S.~Sasaki, M.~Kriener, K.~Segawa, K.~Yada, Y.~Tanaka, M.~Sato, and Y.~Ando:
  Phys. Rev. Lett. {\bfseries 107} (2011) 217001.

\bibitem{Hu}
C.~R. Hu: Phys. Rev. Lett. {\bfseries 72} (1994) 1526.

\bibitem{Tanaka95}
Y.~Tanaka and S.~Kashiwaya: Phys. Rev. Lett. {\bfseries 74} (1995) 3451.

\bibitem{Kashiwaya00}
S.~Kashiwaya and Y.~Tanaka: Rep. Prog. Phys. {\bfseries 63} (2000) 1641.

\bibitem{HaoLee}
L.~Hao and T.~K. Lee: Phys. Rev. B {\bfseries 83} (2011) 134516.

\bibitem{Hsieh}
T.~H. Hsieh and L.~Fu: Phys. Rev. Lett. {\bfseries 108} (2012) 107005.

\bibitem{Koren1}
G.~{Koren}, T.~{Kirzhner}, E.~{Lahoud}, K.~B. {Chashka}, and A.~{Kanigel}:
  Phys. Rev. B {\bfseries 84} (2011) 224521.

\bibitem{Koren2}
G.~Koren and T.~Kirzhner: Phys. Rev. B {\bfseries 86} (2012) 144508.

\bibitem{Stroscio}
N.~{Levy}, T.~{Zhang}, J.~{Ha}, F.~{Sharifi}, A.~A. {Talin}, Y.~{Kuk}, and
  J.~A. {Stroscio}: arXiv:1211.0267 .

\bibitem{Wray1}
L.~A. Wray, S.-Y. Xu, Y.~Xia, D.~Qian, A.~V. Fedorov, H.~Lin, A.~Bansil, L.~Fu,
  Y.~S. Hor, R.~J. Cava, and M.~Z. Hasan: Nat. Phys. {\bfseries 6} (2010) 855.

\bibitem{Wray2}
L.~A. Wray, S.~Xu, Y.~Xia, D.~Qian, A.~V. Fedorov, H.~Lin, A.~Bansil, L.~Fu,
  Y.~S. Hor, R.~J. Cava, and M.~Z. Hasan: Phys. Rev. B {\bfseries 83} (2011)
  224516.

\bibitem{Kriener1}
M.~Kriener, K.~Segawa, Z.~Ren, S.~Sasaki, and Y.~Ando: Phys. Rev. Lett.
  {\bfseries 106} (2011) 127004.

\bibitem{Kriener2}
M.~Kriener, K.~Segawa, Z.~Ren, S.~Sasaki, S.~Wada, S.~Kuwabata, and Y.~Ando:
  Phys. Rev. B {\bfseries 84} (2011) 054513.

\bibitem{Das}
P.~Das, Y.~Suzuki, M.~Tachiki, and K.~Kadowaki: Phys. Rev. B {\bfseries 83}
  (2011) 220513.

\bibitem{Kriener4}
M.~Kriener, K.~Segawa, S.~Sasaki, and Y.~Ando: Phys. Rev. B {\bfseries 86}
  (2012) 180505.

\bibitem{Nagai}
Y.~Nagai, H.~Nakamura, and M.~Machida: Phys. Rev. B {\bfseries 86} (2012)
  094507.

\bibitem{Scalapino}
D.~J. Scalapino: Phys.Rep. {\bfseries 250} (1995) 329.

\bibitem{RevModPhys.67.503}
M.~Sigrist and T.~M. Rice: Rev. Mod. Phys. {\bfseries 67} (1995) 503.

\bibitem{RevModPhys.72.969}
C.~C. Tsuei and J.~R. Kirtley: Rev. Mod. Phys. {\bfseries 72} (2000) 969.

\bibitem{NomuraYamada}
T.~Nomura and K.~Yamada: J. Phys. Soc. Jpn. {\bfseries 71} (2002) 404.

\bibitem{PhysRevLett.87.057001}
M.~E. Zhitomirsky and T.~M. Rice: Phys. Rev. Lett. {\bfseries 87} (2001)
  057001.

\bibitem{PhysRevLett.80.3129}
H.~Tou, Y.~Kitaoka, K.~Ishida, K.~Asayama, N.~Kimura, Y.~Onuki, E.~Yamamoto,
  Y.~Haga, and K.~Maezawa: Phys. Rev. Lett. {\bfseries 80} (1998) 3129.

\bibitem{PhysRevB.77.184515}
K.~Machida and M.~Ichioka: Phys. Rev. B {\bfseries 77} (2008) 184515.

\bibitem{FuBerg}
L.~Fu and E.~Berg: Phys. Rev. Lett. {\bfseries 105} (2010) 097001.

\bibitem{Zhang}
H.~Zhang, C.-X. Liu, X.-L. Qi, X.~Dai, Z.~Fang, and S.-C. Zhang: Nature Phys.
  {\bfseries 5} (2009) 438.

\bibitem{Liu}
C.-X. Liu, X.-L. Qi, H.~Zhang, X.~Dai, Z.~Fang, and S.-C. Zhang: Phys. Rev. B
  {\bfseries 82} (2010) 045122.

\bibitem{Padamsee}
H.~Padamsee, J.~E. Neighbor, and C.~A. Shiffman: J. Low Temp, Phys. {\bfseries
  12} (1973) 387.

\bibitem{BM}
B.~M$\ddot{\mbox u}$hlschlegel: Z. Phys. {\bfseries 155} (1959) 313.

\bibitem{Maruyama}
D.~Maruyama, M.~Sigrist, and Y.~Yanase: J. Phys. Soc. Jpn. {\bfseries 81}
  (2012) 034702.

\end{thebibliography}

\end{document}